\RequirePackage{fix-cm}

\documentclass[smallextended]{svjour3}

\smartqed  

\usepackage{graphicx}
\usepackage{mathptmx}

\begin{document}

\title{A Social Network of Russian ``Kompromat''}
\author{Dmitry Zinoviev}
\institute{D. Zinoviev \at Suffolk University, Boston MA 02114, USA,\\
  Tel.: +1-617-305-1985\\
  ORCID: 0000-0002-8008-0893\\
\email{dzinoviev@suffolk.edu}}
\date{Received: date / Accepted: date}

\maketitle              

\begin{abstract}
  ``Kompromat'' (the Russian word for ``compromising material'') has
  been efficiently used to harass Russian political and business
  elites since the days of the USSR. Online crowdsourcing projects
  such as ``RuCompromat'' made it possible to catalog and analyze
  kompromat using quantitative techniques---namely, social network
  analysis. In this paper, we constructed a social network of 11,000
  Russian and foreign nationals affected by kompromat in Russia in
  1991--2020. The network has an excellent modular structure with 62
  dense communities. One community contains prominent American
  officials, politicians, and entrepreneurs (including President
  Donald Trump) and appears to concern Russia's controversial
  interference in the 2016 U.S. presidential elections. Various
  network centrality measures identify seventeen most central
  kompromat figures, with President Vladimir Putin solidly at the
  top. We further reveal four types of communities dominated by
  entrepreneurs, politicians, bankers, and law enforcement officials
  (``siloviks''), the latter disjointed from the first three.
  \end{abstract}

\section{Introduction}
\label{intro}

``Kompromat'' is a Russian word for ``compromising material.''
Kompromat has been efficiently used to harass political and business
elites. It is widely considered to be essential in shoring up
authoritarian durability~\cite{markowitz2017}. Kompromat regimes often
appear in states with low fiscal capacity and very high police
capacity and harbor widespread criminality combined with systematic
blackmail~\cite{choy2020}.

The practice of kompromat is nothing new: it dates back to the Soviet
period (Stalin, Beria, and the KGB)~\cite{ledeneva2006}. However,
social media advances provided massive and affordable opportunities
for authoritarian regimes to use kompromat against the opposition and
competing factions~\cite{pearce2015}. Simultaneously, online
crowdsourcing projects such as ``RuCompromat''~\cite{rucompromat_com}
make it possible to catalog and systematize kompromat, make it broadly
available to researchers, and potentially disarm it by exposing its
sources.

In this paper, we explore the most extensive online collection of the
post-Soviet (mostly Russian) kompromat hosted by the ``RuCompromat''
team from the perspective of social network analysis---the process of
modeling a social environment as a pattern of regularities in
relationships among interacting units~\cite{wasserman1994}. In our
case, the interacting units are the persons subject to kompromat or
related to kompromat in any other way (e.g., reporters and
politicians~\cite{iakhnis2019networks}). Their co-involvement in
compromising scenarios defines relationships among them. We identify
33 dense groups (network communities) of actors: politicians,
entrepreneurs, and law enforcement officials, in the first
place---associated with specific kompromat cases. Eventually, we
classify the cases in four basic types that differ in the principal
actors' participation.

The rest of the paper is organized as follows: In
Section~\ref{sec:data}, we describe the data set, its provenance, and
its structure; in Section~\ref{sec:construction}, we explain the
network construction and analyze it; in Section~\ref{sec:discussion},
we present the results and discuss them. Finally, in
Section~\ref{sec:conclusion}, we conclude and lay the ground for
further work.

\section{\label{sec:data}Dataset}

We collected the dataset in September 2020 from the site
``RuCompromat'' \cite{rucompromat_com}. The site provides information
about approximately 11 thousand persons, both Russian and foreign
nationals. For each person, the cite provides references to media
articles (only for the years 2013--2020) that mention potentially
compromising facts about the person, a list of other people mentioned
together with the person, and a list of organizations (e.g.,
companies, banks, and government offices) related to the person.

As an example, an article called ``Patrushev Jr. and the Orthodox
retirees. Unprofitable Rosselkhozbank goes to the aid of ``Peresvet,''
published by the ``Ruspress'' agency~\cite{rospres2016}, mentions
banker Valery Meshalkin and the son of the former FSB director Nikolai
Patryshev. Thus, ``RuCompromat'' lists the two as related.

The dataset contains co-references for 11,118 persons for the time
frame from 1999 to 2020. We have not validated the references' accuracy
and take the ``RuCompromat'' contributors' opinions as the ground
truth.

\section{\label{sec:construction}Network Construction and Analysis}
\begin{figure}
\includegraphics[width=\textwidth]{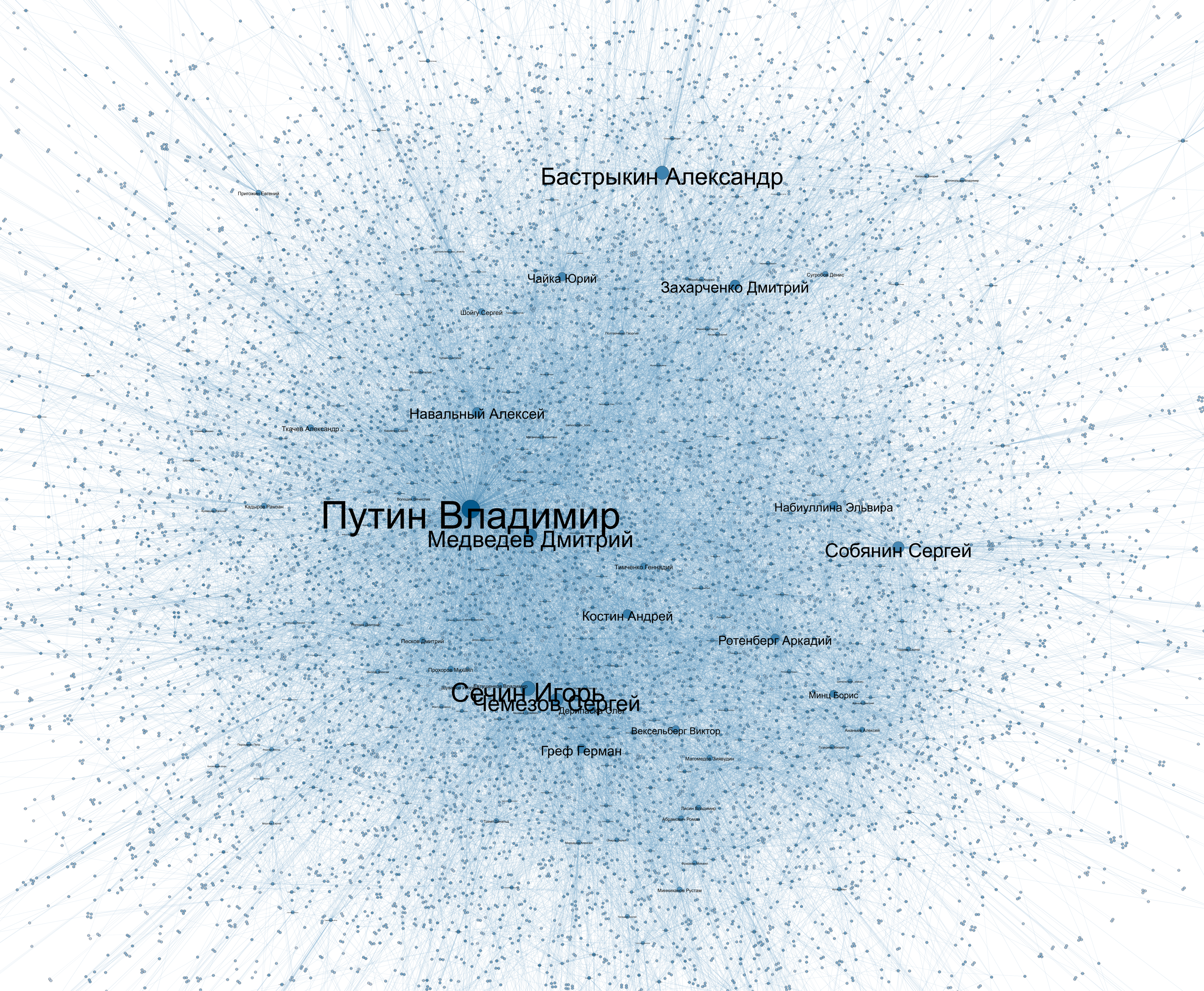}
\caption{\label{fig:core}The dense core of the ``kompromat'' network,
  with Vladimir Putin and Dmitry Medvedev at the center, Igor Sechin
  and Sergey Chemezov below, Sergey Sobyanin on the right, and
  Aleksandr Bastrykin above.}
\end{figure}

The first step in social network analysis is the construction of a
social network. The network consists of 11,118 nodes representing
persons, and 37,544 edges representing relationships between the
persons. The density of the network is $\approx\!0.0006$. The network
is unweighted (all edges have the same weight of 1), undirected (if
person A is connected to person B, then B is also connected to A), and
strongly connected (it is possible to get from any person A to any
person B by following edges without leaving the network). The diameter
of the network is 12. Fig.~\ref{fig:core} shows the central part of
the network---its dense core.

\begin{figure}
\includegraphics[width=\textwidth]{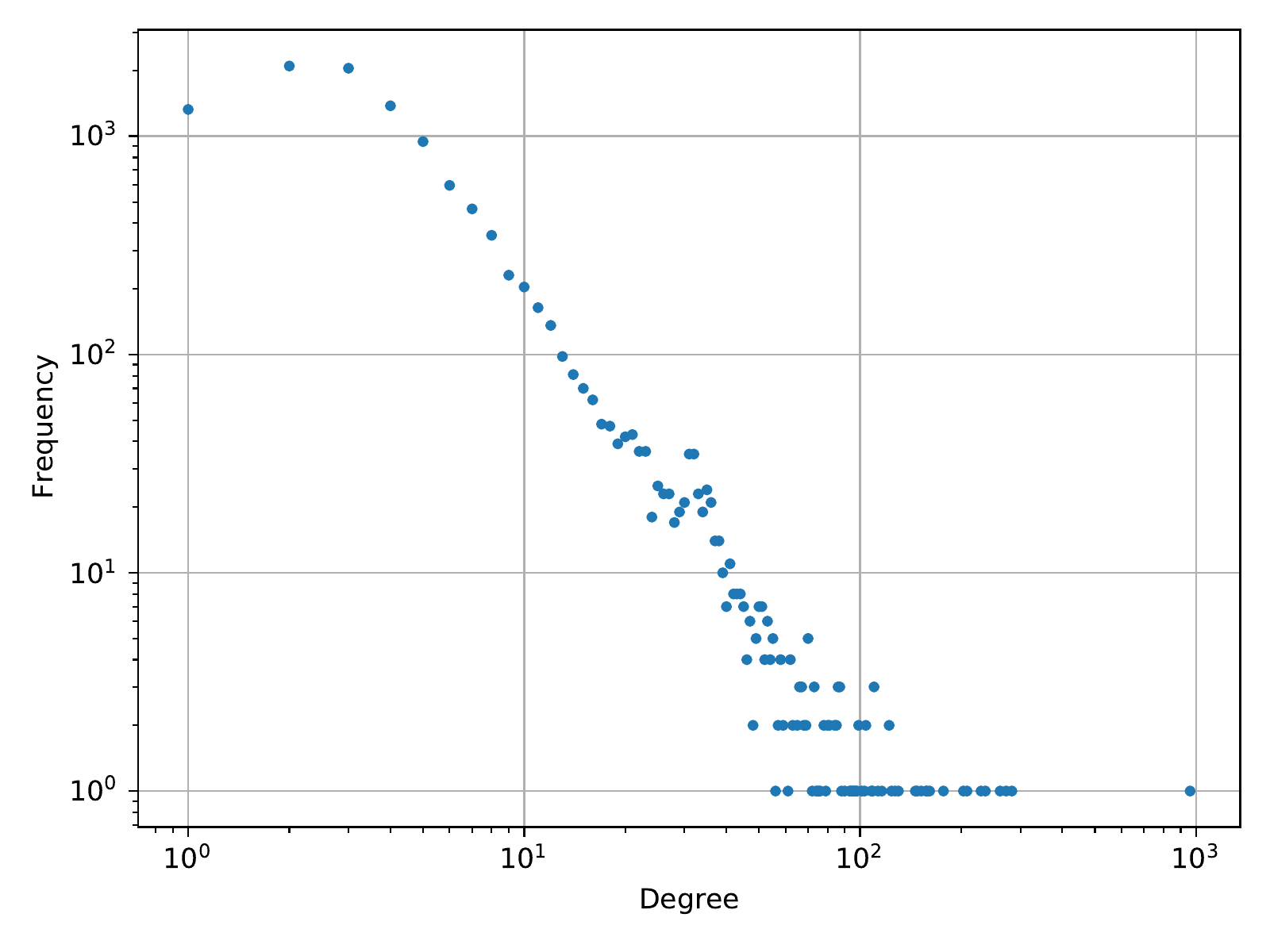}
\caption{\label{fig:degree}The node degree distribution has a ``long
  tail'' and can be approximated by the power law $f_d\sim d^{-1.85}$
  for $d\ge3$.}
\end{figure}

Fig.~\ref{fig:degree} shows the node degree distribution in the
network. The distribution has a ``long tail'' and can be approximated
by the power law with the exponent $\alpha\approx\!-1.85$ (for the
nodes with three or more neighbors). This exponent is typical for
massive online social networks~\cite{shi2007}. The majority of persons
are connected poorly, only to one (2,100) or two (2,051) other
persons. On the other end of the spectrum, there are the
highest-connected (most embedded) persons: Vladimir Putin (President
of Russia; 960 connections), Igor Sechin (CEO of Rosneft', the Russian
state oil company; 283), Dmitry Medvedev (former President and Prime
Minister; 272), Sergey Chemezov (CEO of Rostec Corporation; 261),
Sergey Sobyanin (Mayor of the City of Moscow; 236), Aleksandr
Bastrykin (head of the Investigative Committee; 220), Aleksey Navalny
(opposition leader; 208), and Arkady Rotenberg (co-owner of
Stroygazmontazh construction group; 203)---{\em cf.}
Fig.~\ref{fig:core} and Table~\ref{table:top10}.

We used the Louvain community detection algorithm~\cite{blondel08} to
partition the network into 62 network communities: groups of persons
that are more closely connected to each other than to the persons from
the rest of the network. Each of the communities is denser than the
whole network (Table~\ref{table:communities}). The sizes of the
communities range from 3 persons to 1,470 persons. The partition
modularity~\cite{newman2006} is 0.64 on the scale from -0.5 (no
community structure) to 1.0 (perfect community structure).

As it is customary in social network analysis, we excluded the smallest
29 communities with fewer than 100 persons from the study. For each node
in the remaining communities, we calculated the clustering coefficient and
several centralities: degree centrality, closeness centrality,
betweenness centrality, and eigenvector centrality~\cite{freeman1979}.

The degree centrality, or simply the degree of a node, is the number
of the node's connections. A person with a high degree centrality has
been affected by more kompromat cases.

To understand closeness centrality, let us consider two randomly
chosen persons: Vladimir Putin and Zhou Yongkang (a late senior leader
of the Communist Party of China, CPC). Due to the network's sparse
nature, it is improbable that they shared a direct connecting
edge. Putin is more likely connected to another person: say, Xi
Jinping (the General Secretary of the CPC)---who, consequently, was
connected to Zhou. We say that Putin is two hops apart from Zhou.

In a different scenario, Putin is connected to Genry Reznik (a
prominent Russian lawyer), who is connected to Yury Antonov (a retired
activist from St.~Petersburg), who is connected to Inna Pashchenko (a
WWII veteran; all three are actual members of the ``RuCompromat''
dataset). Therefore, Putin is three hops apart from Pashchenko.

There is a path from Putin to the other person in both scenarios, and
the length of the path is two and three, respectively. In general, the
average length of the shortest paths from a person to all other people
in the network is called the closeness centrality of that person. A
person with a high closeness centrality has been more likely
indirectly affected by or involved in more kompromat-related cases.

Quite expectedly~\cite{valente2008}, the degree centrality is
positively correlated with closeness centrality (their correlation is
$r\approx0.867$). On these grounds, we excluded the degree centrality
from the results.

Any person along a path is somewhat similar to any other person along
the same path. The shorter is the distance between two such persons,
the more similar they are, to the extent that two immediate neighbors
share the same kompromat. The number of the shortest paths that pass
through a person is called the betweenness centrality of the person. A
person with a high betweenness centrality is similar to more persons
along the adjacent paths---in other words, that person is more
typical. Thus, betweenness centrality appears to be a sensible measure
of typicality.

Let us assume that each person has a quantitative ``kompromat score''
that somehow measures the extent to which the person was subject to
kompromat. Let us also assume that a compromising document ``smears''
all referenced persons proportionally to the kompromat scores of
everyone involved in the case. In other words, being mentioned
together with a rapist and a thief paints a person as partially a
rapist and partially a thief, regardless of the actual case context:
''Tell me who your friends are, and we will tell you who you are.''
Such a score is called the eigenvector centrality. The eigenvector
centrality measures a person's influence on the network---in this
study's context, the person's toxicity.

The last network attribute is the clustering
coefficient~\cite{watts1998}. It measures how many immediate network
neighbors of a person are also directly connected. A high clustering
coefficient is a sign of a tightly knit group. In a fully-connected
clique, the clustering coefficient is 1. In a ``hub-and-spoke''
network, it is 0.

\begin{table}
  \caption{\label{table:communities}Mean community parameters, sorted
    in the decreasing order of the Betweenness centrality: Size
    (number of persons), Closeness centrality, Eigenvector centrality,
    Clustering Coefficient, and Density. The mean betweenness
    centrality of the top six communities is above the network
    average.}
    \begin{tabular}{rlrrrrrrr}
      \hline\noalign{\smallskip}
      &Community label         & B    & S     & C    & E     & CC & D        \\\noalign{\smallskip}\hline\noalign{\smallskip}
      1&Putin Vladimir       & 0.54 & 1,470 & 0.24 &  7.80  & 0.53 & 0.0124  \\
      2&Bastrykin Aleksandr  & 0.39 &  749  & 0.23 &  3.44  & 0.61 & 0.0065  \\
      3&Sobyanin Sergey      & 0.38 &  940  & 0.23 &  4.00  & 0.56 & 0.0208  \\
      4&Sechin Igor          & 0.35 &  199  & 0.23 &  2.93  & 0.58 & 0.0124  \\
      5&Rotenberg Arkady     & 0.35 &  382  & 0.23 &  3.99  & 0.57 & 0.0035  \\
      6&Kostin Andrey        & 0.33 &  233  & 0.23 &  3.30  & 0.64 & 0.0075  \\\noalign{\smallskip}\hline\noalign{\smallskip}
      7&Chemezov Sergey      & 0.31 &  322  & 0.23 &  3.83  & 0.59 & 0.0206  \\
      8&Nabiullina Elvira    & 0.30 &  390  & 0.22 &  2.70  & 0.58 & 0.0144  \\
      9&Shuvalov Igor        & 0.29 &  176  & 0.22 &  1.97  & 0.62 & 0.0152  \\
      10&Luzhkov Yury        & 0.29 &  298  & 0.22 &  2.12  & 0.59 & 0.0309  \\
      11&Rogozin Dmitry      & 0.28 &  293  & 0.22 &  3.11  & 0.59 & 0.0228  \\
      12&Poltavchenko Georgy & 0.28 &  230  & 0.21 &  2.43  & 0.61 & 0.0118  \\
      13&Gref German         & 0.27 &  353  & 0.22 &  2.06  & 0.61 & 0.0123  \\
      14&Kerimov Suleyman    & 0.27 &  225  & 0.23 &  2.65  & 0.63 & 0.0341  \\
      15&Chayka Yury         & 0.27 &  338  & 0.22 &  2.75  & 0.60 & 0.0451  \\
      16&Mutko Vitaly        & 0.27 &  121  & 0.22 &  2.09  & 0.63 & 0.0129  \\
      17&Abramovich Roman    & 0.26 &  465  & 0.23 &  3.21  & 0.61 & 0.0242  \\
      18&Ismailov Telman     & 0.25 &  105  & 0.21 &  2.34  & 0.63 & 0.0185  \\
      19&Tolokonsky Viktor   & 0.25 &  264  & 0.21 &  2.08  & 0.65 & 0.0108  \\
      20&Poroshenko Petr      & 0.25 &  385  & 0.23 &  2.74  & 0.63 & 0.3636 \\
      21&Trump Donald         & 0.24 &  289  & 0.22 &  2.87  & 0.59 & 0.0130 \\
      22&Kadyrov Ramzan       & 0.24 &  202  & 0.23 &  3.95  & 0.64 & 0.0163 \\
      23&Ivanov Sergey        & 0.24 &  230  & 0.22 &  2.98  & 0.62 & 0.0434 \\
      24&Vekselberg Viktor    & 0.23 &  331  & 0.22 &  1.92  & 0.65 & 0.0154 \\
      25&Khodorkovsky Mikhail & 0.23 &  142  & 0.22 &  3.91  & 0.65 & 0.0139 \\
      26&Avetisyan Artem      & 0.23 &  100  & 0.22 &  2.58  & 0.62 & 0.0151 \\
      27&Miller Aleksey       & 0.23 &  107  & 0.21 &  2.06  & 0.64 & 0.0365 \\
      28&Yakunin Vladimir     & 0.22 &  340  & 0.23 &  4.59  & 0.59 & 0.0174 \\
      29&Chayka Igor          & 0.21 &  307  & 0.23 &  2.92  & 0.61 & 0.0123 \\
      30&Mordashev Aleksey    & 0.21 &  156  & 0.22 &  1.58  & 0.64 & 0.0194 \\
      31&Mikhaylov Sergey     & 0.20 &  182  & 0.22 &  2.18  & 0.61 & 0.0316 \\
      32&Prigozhin Evgeny     & 0.20 &  152  & 0.22 &  1.67  & 0.65 & 0.0235 \\
      33&Radaev Valery        & 0.19 &  130  & 0.22 &  2.01  & 0.66 & 0.0214 \\\noalign{\smallskip}\hline\noalign{\smallskip}
        &Overall              & 0.32 &  n/a  & 0.23 &  3.56  & 0.60 & 0.0006 \\\noalign{\smallskip}\hline
    \end{tabular}
\end{table}

\begin{figure}
\includegraphics[width=\textwidth]{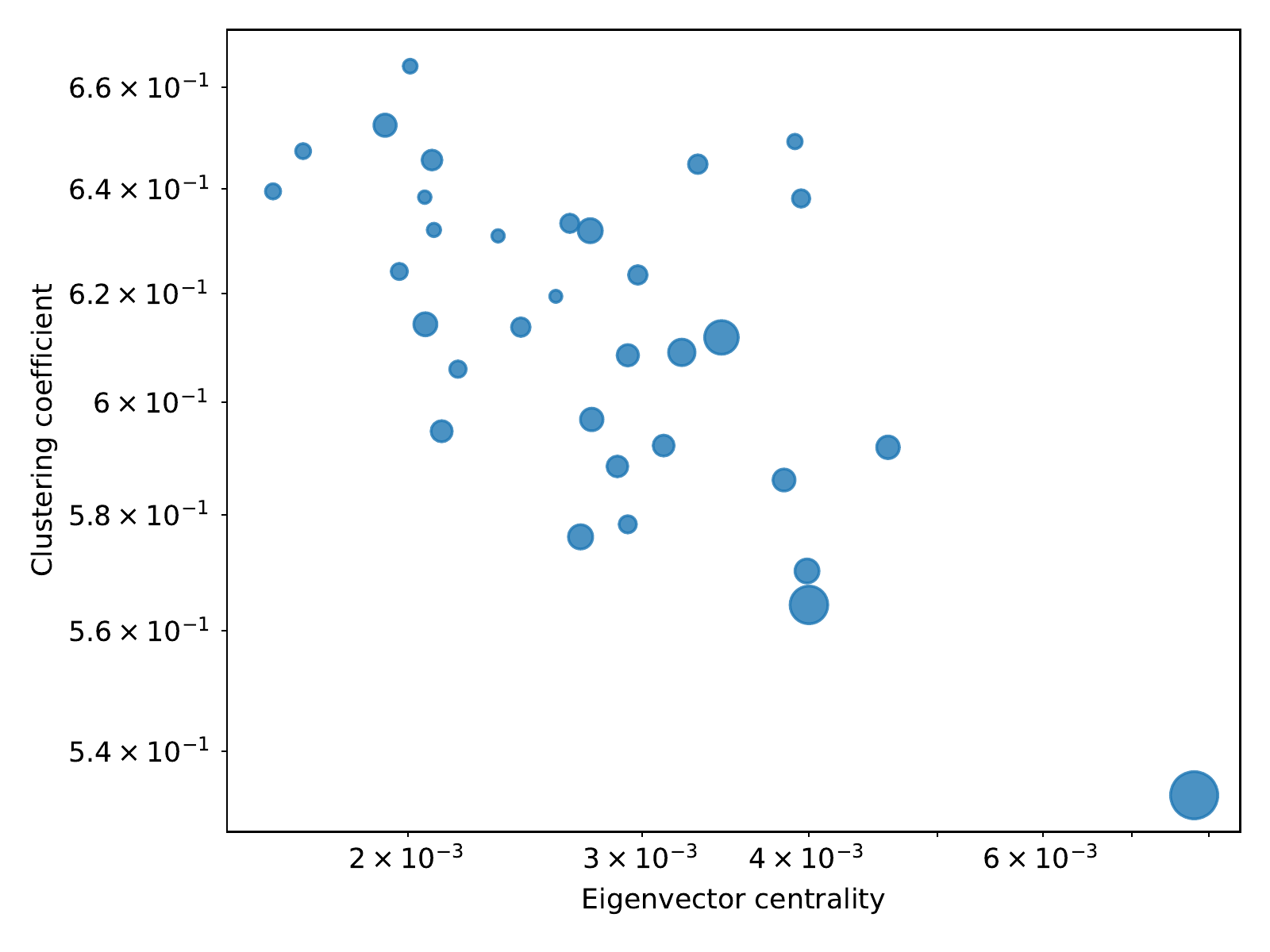}
\caption{\label{fig:eigen_vs_cc}Mean eigenvector centrality and
  clustering coefficient for the major 33 communities (see
  Table~\ref{table:communities}). The size of a dot represents the number
  of persons in the community.}
\end{figure}

Table~\ref{table:communities} shows the mean values of each
community's network attributes, sorted in the decreasing order of the
betweenness centrality, and the number of persons in the
communities. Note that the mean closeness centrality in all
communities is almost the same. On the contrary, the clustering
coefficient is negatively correlated with the betweenness and
eigenvector centralities (Fig.~\ref{fig:eigen_vs_cc}). The dense
communities (higher clustering coefficient) are, on average, smaller,
less typical/more peripheral (lower betweenness centrality), and less
toxic (lower eigenvector centrality).

\begin{table}
  \caption{\label{table:top10}Top ten persons, sorted in the
    decreasing order of the betweenness centrality, closeness
    centrality, eigenvector centrality, and degree. The
    dagger$^\dagger$ marks the persons who appear in more than one
    column.}
  \begin{tabular}{llll}
    \hline\noalign{\smallskip}
    Betweenness& Closeness & Eigenvector& Degree \\\noalign{\smallskip}\hline\noalign{\smallskip}
    (Typicality)& (Indirect involvement) & (Toxicity)& (Direct involvement) \\\noalign{\smallskip}\hline\noalign{\smallskip}
    Putin$^\dagger$ V. & Putin V.$^\dagger$& Putin V.$^\dagger$& Putin V.$^\dagger$\\
    Sechin I.$^\dagger$& Medvedev D.$^\dagger$& Medvedev D.$^\dagger$& Sechin I.$^\dagger$\\
    Bastrykin A.$^\dagger$ & Kostin A.$^\dagger$& Sechin I.$^\dagger$& Medvedev D.$^\dagger$\\
    Mevedev D.$^\dagger$ & Sechin I.$^\dagger$& Chemezov S.$^\dagger$& Chemezov S.$^\dagger$\\
    Chemezov S.$^\dagger$ & Zakharchenko D.$^\dagger$ & Rotenberg A.$^\dagger$&Sobyanin S.$^\dagger$\\
    Sobyanin S.$^\dagger$ & Klimenko G. & Navalny A.$^\dagger$& Bastrykin A.$^\dagger$\\
    Zakharchenko D.$^\dagger$ & Kamenschchik D. & Timchenko G. & Navalny A.$^\dagger$\\
    Navalny A.$^\dagger$ & Mints B. & Bastrykin A.$^\dagger$& Rotenberg A.$^\dagger$\\
    Gref G.$^\dagger$ & Chemezov S.$^\dagger$ & Deripaska O.$^\dagger$& Deripaska O.$^\dagger$\\
    Kostin A.$^\dagger$ & Oreshkin M. & Kostin A.$^\dagger$&Gref G.$^\dagger$\\\noalign{\smallskip}\hline
  \end{tabular}
\end{table}

Table~\ref{table:top10} shows the top ten persons in each centrality
category. Only five persons: German Klimenko (Vladimir Putin's
Internet advisor), Dmitry Kamenshchik (owner of Moscow Domodedovo
airport), Boris Mints (former owner of the Future Financial Group),
Mikhail Oreshkin (the Minister for Economic Development), and Gennady
Timchenko (former co-owner of Gunvor Group Ltd)---appear once. Most of
the top performers shine in several categories: Vladimir Putin (\#1 in
all four categories), Sergey Chemezov (appears in 4 categories),
Dmitry Medvedev (4), Igor Sechin (4), Aleksandr Bastrykin (3), Aleksey
Navalny (3), Oleg Deripaska (the founder of Basic Element, one of
Russia's largest industrial groups; 2), German Gref (CEO of Sberbank;
2), Andrey Kostin (President of VTB Bank; 2), Arkady Rotenberg (2),
Sergey Sobyanin (2), and Dmitry Zakharchenko (a former anti-corruption
official convicted of bribery; 2). They represent almost all principal
groups of kompromat stakeholders: business, politics, law enforcement,
banking, government, and press---with a notable exception of the
criminal underworld.

\section{\label{sec:discussion}Discussion}

In the rest of the paper, we investigate the composition of the
communities, their relationships, and kompromat cases' typology.

\begin{figure}
\includegraphics[width=\columnwidth]{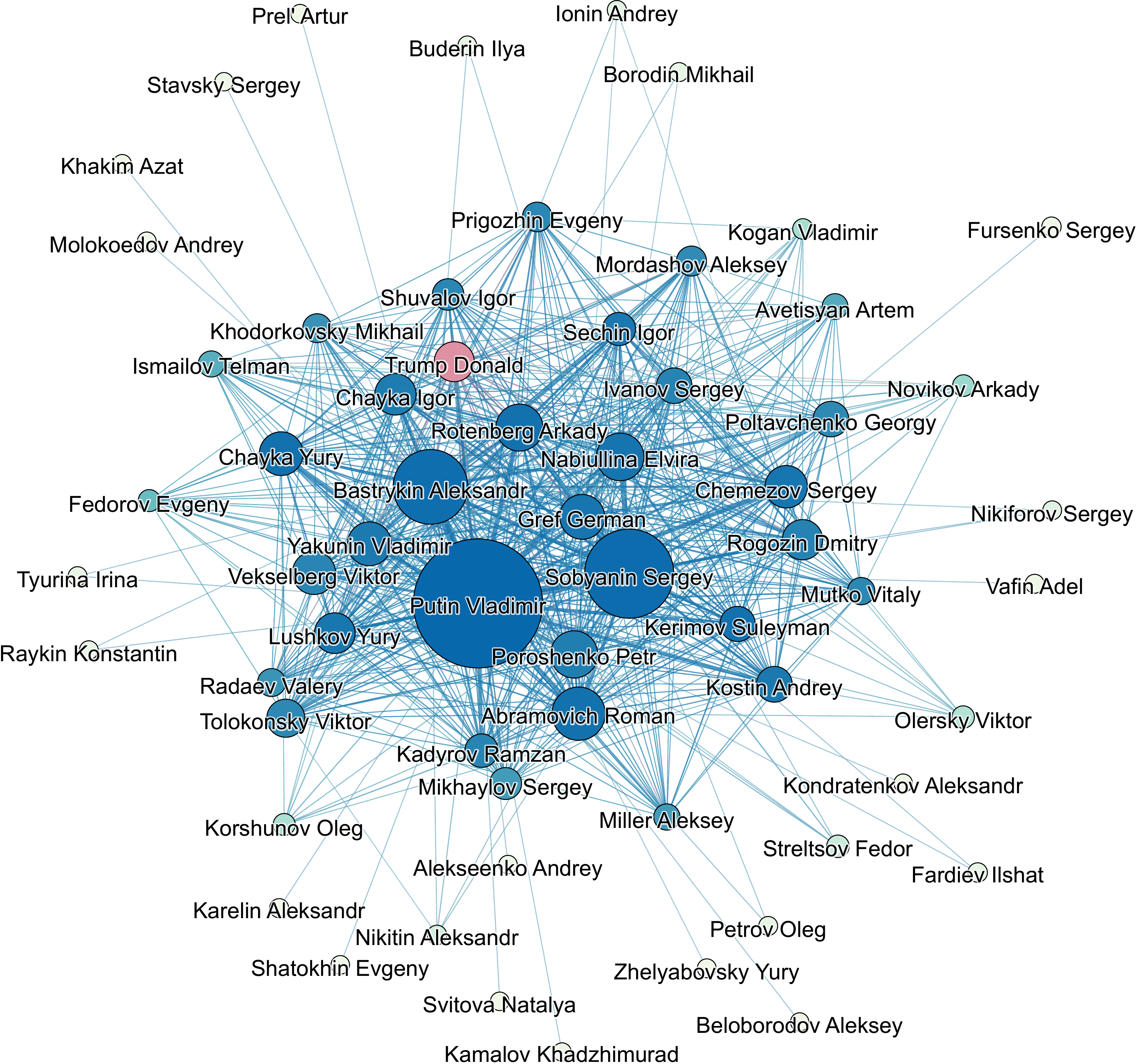}
\caption{\label{fig:induced}An induced kompromat network: a birds-eye
  view of the original network in which the nodes represent communities.}
\end{figure}

Fig.~\ref{fig:induced} presents a birds-eye view of the original
kompromat network, called an induced network. Each node in the induced
network stands for a community in the original network. Node size
represents the number of persons in the community. The induced network
edges are weighted; their weight (thickness) represents the number of
individual edges in the original network. The communities are named after
the most prominent persons---the persons with the highest betweenness
centrality. The node color represents the mean betweenness centrality
of the community nodes: darker nodes have a higher centrality. (One
exception is the node labeled ``Trump Donald''---it is painted pink
because it contains disproportionally many foreign nationals.)

\begin{table}
    \caption{\label{table:top5names}Top five most prominent persons
      (in terms of betweenness centrality) in each community (see also
      Table~\ref{table:communities}). The dagger$^\dagger$ denotes
      English-language duplicates. The star$^\star$ denotes foreign
      nationals. The equal sign$^=$ denotes the same person known
      under two different names.}
    \begin{tabular}{rll}\hline\noalign{\smallskip}
        &Community label &Other persons\\\noalign{\smallskip}\hline\noalign{\smallskip}
       1&Putin V. & Medvedev D., Navalny A., Deripaska O., Timchenko G.\\
       2&Bastrykin A. & Zakharchenko D., Sugrobov D., Titov B., Chubays A.\\
       3&Sobyanin S. & Magomedov Z., Gutseriev M., Mints B., Ananyev A.\\
       4&Sechin I. & Sechina M., Leontyev M., Rakhmanov A., Avdolyan A.\\
       5&Rotenberg A. & Peskov D., Rotenberg B., Lisin V., Rotenberg I.\\
       6&Kostin A. & Evtushenkov V., Rakhimov U., Gagiev A., Khamitov R.\\
       7&Chemezov S. & Prokhorov M., Manturov D., Khudaynatov E., Ignatova E.\\
       8&Nabiullina E. & Tokarev N., Bedzhamov G., Pugachev S., Markus L.\\
       9&Shuvalov I. & Shuvalov I.$^\dagger$, Kesaev I., Levchenko S., Filev V.\\
      10&Luzhkov Yu. & Chayka A., Yurevich M., Dubrovsky B., Rashnikov V.\\
      11&Rogozin D. & Lavrov S., Patrushev N., Komarov I., Vinokurov A.\\
      12&Poltavchenko G. & Beglov A., Albin I.$^=$, Slyunyaev I.$^=$, Oganesyan M.\\
      13&Gref G. & Usmanov A., Mamut A., Durov P., Dmitriev V.\\
      14&Kerimov S. & Matvienko V., Turchak A., Golodets O., Petrenko S.\\
      15&Chayka Yu. & Shoygu S., Serdyukov A., Vasilyeva E., Ivanov T.\\
      16&Mutko V. & Rashkin V., Ablyazov M.$^\star$, Rodchenkov G., McLaren R.$^\star$\\
      17&Abramovich R. & Vorobyev A., Potanin V., Sobchak K., Lebedev V.\\
      18&Ismailov T. & Usoyan A., Mitrofanov A., Dzhaniev R., Varshavsky A.\\
      19&Tolokonsky V. & Morozov S., Shantsev V., Khinshteyn A., Nazarov V.\\
      20&Poroshenko P.$^\star$ & Yanukovich V.$^\star$, Surkov V., Belykh N., Kolomoysky I.$^\star$\\
      21&Trump D.$^\star$ & Cherkalin K., Mishustin M., Belousov A., Tkachev I.\\
      22&Kadyrov R. & Ulyukaev A., Nemtsov B., Timakova N., Galchev F.\\
      23&Ivanov S. & Gordeev A., Skrynnik E., Ivanov A., Korolev O.\\
      24&Vekselberg V. & Golunov I., Golubev V., Varshavsky V., Blavatnik L.$^\star$\\
      25&Khodorkovsky M. & Skripal S., Petrov A., Litvinenko V., Lebedev P.\\
      26&Avetisyan A. & Calvi M.$^\star$, Nazarbaev N.$^\star$, Maduro N.$^\star$, Delpal P.$^\star$\\
      27&Miller A. & Slipenchuk M., Khlebnikov P.$^\star$, Lurakhmaev V., Lanin M.\\
      28&Yakunin V. & Belozerov O., Tikhonova E., Tikhonova K.$^\dagger$, Gorbuntsov G.\\
      29&Chayka I. & Traber I.$^\star$, Zhirinovsky V., Skoch A., Yarovaya I.\\
      30&Mordashov A. & Shvets A., Novak A., Cyril (Patriarch), Khotimsky S.\\
      31&Mikhaylov S. & Malofeev K., Strelkov I., Kaspersky E., Girkin I.\\
      32&Prigozhin E. & Prigozhin E.$^\dagger$, Kligman I., Gerasimenko A., Uss A.\\
      33&Radaev V. & Savelyev V., Lebedev A., Shishov O., Vantsev V.\\\noalign{\smallskip}\hline
\end{tabular}
\end{table}

Table~\ref{table:top5names} additionally shows the top five most
prominent persons in each community (some names have been spelled in
Russian and English in the original dataset, which resulted in
duplications; also, Igor Slyunyaev changed his unpleasantly sounding
name to Igor Albin and was recorded in the same community under two
names).
    
The community \#21, labeled ``Trump Donald,'' stands out among other
communities. Not only its most prominent representative is a foreign
national and the president of a sovereign nation, but the nation---the
USA---is also well outside of Russia's zone of influence. At a closer
look, the community contains other prominent American officials,
politicians, and entrepreneurs such as businessmen Ilon Mask, Bill
Gates, and Jeff Bezos, U.S. government officials Robert Mueller and
Rex Tillerson, and a disgraced lawyer Michael Cohen. The community is
also the home to the Russian prime minister Mikhail Mishustin and
Maria Butina, who pleaded guilty to conspiracy to act as an
unregistered foreign agent of Russia within the
USA~\cite{swaine2018maria}. The community's composition suggests that
it concerns Russia's controversial interference in the 2016
U.S. presidential elections~\cite{hayes2017trump}.

In the final stage of the analysis, we identified several types of
communities, based on the affiliations of their most prominent members
with one of the following categories: ``business,'' including state
corporations (53 people in Table~\ref{table:top5names}), ``politics''
(50), ``law enforcement'' (known in Russia as ``siloviks,'' or
``people of force''~\cite{reddaway20182007}; 16), ``banking'' (15),
``government officials'' (13), ``criminal world'' (6), ``press'' (5),
and ``others'' (4). Sometimes, there was more than one affiliation per
person: for example, Igor Sechin, as the CEO of Rosneft', is an
entrepreneur, but since Rosneft' is a state oil company, he is also a
government official. In such cases, we selected the most notable
affiliation.

As a result, we described each community with eight numbers---in other
words, represented it as a point in 8-dimensional space, to a total of
33 points. For example, of the five most prominent persons in
community \#1, two are entrepreneurs, two are politicians, and one is
as journalist. The numbers for that community are (2, 2, 0, 0, 0, 0,
1, 0).

Such multi-dimensional points could be arranged into groups by
applying k-means clustering---a method that aims to partition the
observations into k clusters in which each observation belongs to the
cluster with the nearest mean. We chose $k=4$ in the expectation of a
match with contemporary Russian kompromat's typology that includes
political, economic, criminal, and personal types of incriminating
material~\cite{ledeneva2006}. Table~\ref{table:types} presents the
results of the classification. Each network community and, by
inclusion, all individual members of those communities, belong to one
of the four types.

\begin{table}
  \caption{\label{table:types}Kompromat types: ``business''
    T$_1$, ``politics'' T$_2$, ``banking'' T$_3$, and ``law
    enforcement'' T$_4$.}
\begin{tabular}{lrrrr}
\hline\noalign{\smallskip}
Type          &  T$_1$        & T$_2$     &  T$_3$      & T$_4$\\\noalign{\smallskip}\hline\noalign{\smallskip}
Business      &  \underline{39}  &       7&        4 & 3 \\
Politics      &  11        &\underline{29}&        8 & 2 \\
Banking       &  4         &    --- & \underline{11} & ---  \\
Law Enforcement& 1         &       2&        2 & \underline{11}\\
Government    &  5         &       2&        3 & 3\\
Criminal      &  6         &    --- &       ---& ---\\
Press         &  5         &    --- &       ---& ---\\
Other         &  2         &    --- &        1 & 1\\\noalign{\smallskip}\hline\noalign{\smallskip}
\# of communities          &  15        &       8&        6 & 4 \\\noalign{\smallskip}\hline
\end{tabular}
\end{table}

The results, while different from the taxonomy proposed
in~\cite{ledeneva2006}, appear consistent. Each kompromat type has the
specific, dominant category of participants: entrepreneurs (T$_1$),
politicians (T$_2$), bankers (T$_3$), and ``siloviks''
(T$_4$). However, types T$_1$ and T$_3$ also have a significant
secondary population of politicians, and T$_2$ additionally includes
entrepreneurs. Thus, the first three types represent a corrupted
symbiosis of industrial and banking capital and political power,
biased towards one of the factions, depending on the type.

The fourth type, T$_4$, comprises the ``siloviks'' and has few
representatives from the other categories. The difference suggests
that the kompromat cases involving law enforcement officials, though
not entirely isolated, differ from those that have affected the
political and economic block.

\section{\label{sec:conclusion}Conclusion and Future Work}

In this paper, we analyzed a social network of 11,000 Russian and
foreign nationals, including politicians, entrepreneurs, bankers, law
enforcement officials, and high-profile criminals, affected by
kompromat: compromising materials. The data for the study was
downloaded from ``RuCompromat,'' a Russian online encyclopedia of
kompromat. The network is modular and has an excellent community
structure. Each network community brings together persons who
participated in similar kompromat cases. We calculated the network
attributes (such as various centralities and clustering coefficient)
and identified the most prominent persons in the whole network and
each community. By looking at the top community members' affiliations
with various socio-economic groups, we introduced a four-type taxonomy
of the communities. Three types represent industrial and banking
capital and political power; law enforcement officials (the
``siloviks'') dominate the fourth.

``RuCompromat'' offers an additional level of information:
organizations involved in the kompromat cases. In the future, this
information can be used to construct and analyze a joint
socio-organizational network. Since organizations are usually easier
to classify than individuals, adding them to the dataset could help us
automatically assign persons to the categories, which would improve
the kompromat cases' typology.

\section*{Acknowledgment}

The author is grateful to Pelin Bi\c{c}en, Professor of Marketing at
Suffolk University, and Vasily Gatov, Russian media researcher and
author, for their encouragement and helpful suggestions.

\section*{Conflict of interest}
The authors declare that they have no conflict of interest.
 
\section*{Funding}
Not applicable.

\section*{Availability of data and material}
Not applicable.

\section*{Code availability}
Not applicable.

\bibliographystyle{spmpsci}
\bibliography{cs}

\begin{thebibliography}{10}
\providecommand{\url}[1]{{#1}}
\providecommand{\urlprefix}{URL }
\expandafter\ifx\csname urlstyle\endcsname\relax
  \providecommand{\doi}[1]{DOI~\discretionary{}{}{}#1}\else
  \providecommand{\doi}{DOI~\discretionary{}{}{}\begingroup
  \urlstyle{rm}\Url}\fi

\bibitem{blondel08}
Blondel, V., Guillaume, J.L., Lambiotte, R., Lefebvre, E.: {Fast Unfolding of
  Communities in Large Networks}.
\newblock J. of Statistical Mechanics: Theory and Experiment (10), 1000 (2008)

\bibitem{choy2020}
Choy, J.: Kompromat: A theory of blackmail as a system of governance.
\newblock Journal of Development Economics p. 102535 (2020)

\bibitem{freeman1979}
Freeman, L.: {Centrality in Social Networks: Conceptual Clarification}.
\newblock Social Networks \textbf{1}, 215--239 (1979)

\bibitem{hayes2017trump}
Hayes, N.: Trump's kompromat  (2017)

\bibitem{iakhnis2019networks}
Iakhnis, E., Badawy, A.: Networks of power: Analyzing world leaders
  interactions on social media.
\newblock arXiv preprint arXiv:1907.11283 (2019)

\bibitem{ledeneva2006}
Ledeneva, A.V.: How {R}ussia really works: The informal practices that shaped
  post-{S}oviet politics and business.
\newblock Cornell University Press (2006)

\bibitem{markowitz2017}
Markowitz, L.: Beyond kompromat: Coercion, corruption, and deterred defection
  in {U}zbekistan.
\newblock Comparative Politics \textbf{50}(1), 103--121 (2017)

\bibitem{newman2006}
Newman, M.: Modularity and community structure in networks.
\newblock Proceedings of the National Academy of Sciences of the United States
  of America \textbf{103}(23), 8577--8696 (2006)

\bibitem{pearce2015}
Pearce, K.: Democratizing kompromat: the affordances of social media for
  state-sponsored harassment.
\newblock Information, Communication \& Society \textbf{18}(10), 1158--1174
  (2015)

\bibitem{reddaway20182007}
Reddaway, P.: 2007--2008: War fades, tandem forms, {C}herkesov clan dissolves,
  {S}echinites decline; {P}utin generates new factional wars—general
  procuracy vs the investigations committee of the procuracy (icp), {M}edvedev
  vs {S}echin.
\newblock In: Russia’s Domestic Security Wars, pp. 89--93. Springer (2018)

\bibitem{rucompromat_com}
{RuCompromat: Encyclopedia of the Library of Kompromat}.
\newblock Retrieved in 2020.
\newblock \urlprefix\url{http://rucompromat.com/}

\bibitem{rospres2016}
Ruspress: Patryshev-mladshiy i pravoslavnye pensionery. {U}bytochnyy
  {R}os\-sel'\-khozbank idyot na pomoshch ``{P}eresvetu''.
\newblock Published online in Russian (2018).
\newblock \urlprefix\url{http://www.rospres.com/finance/18391/}.
\newblock Removed, cached copy available

\bibitem{shi2007}
Shi, X., Adamic, L., Strauss, M.: {Networks of strong ties}.
\newblock Physica A: Statistical Mechanics and its Applications
  \textbf{378}(1), 33--47 (2007).
\newblock \doi{10.1016/j.physa.2006.11.072}

\bibitem{swaine2018maria}
Swaine, J.: Maria {B}utina’s alleged backer linked to {K}remlin-financed bank
  and {P}utin associates.
\newblock The Guardian (2018).
\newblock
  \urlprefix\url{https://www.theguardian.com/world/2018/aug/06/\-maria-butina-charged-spying-putin-russiakremlin}

\bibitem{valente2008}
Valente, T., Coronges, K., Lakon, C., Costenbader, E.: How correlated are
  network centrality measures?
\newblock Connections (Toronto, Ont.) \textbf{28}(1), 16 (2008)

\bibitem{wasserman1994}
Wasserman, S., Faust, K.: Social network analysis: Methods and applications,
  vol.~8.
\newblock Cambridge university press (1994)

\bibitem{watts1998}
Watts, D.J., Strogatz, S.H.: Collective dynamics of ``small-world'' networks.
\newblock Nature \textbf{393}(6684), 440--442 (1998)

\end{thebibliography}
\end{document}